\documentclass[prc,preprint]{revtex4}

\usepackage{graphicx}
\usepackage{amsmath}
\begin{document}
%\twocolumn[\hsize\textwidth\columnwidth\hsize\csname@twocolumnfalse\endcsname

\title{{\bf Spatially dependent atom-photon entanglement}}

\author{{\bf Zahra Amini Sabegh, Rahim Amiri and Mohammad Mahmoudi \footnote{E-mail: mahmoudi@znu.ac.ir}}}
 \affiliation{Department of Physics, Zanjan University, University Blvd, 45371-38791, Zanjan, Iran}

%\author{ M Mahmoudi, M A Maleki and Z Amini Sabegh }
%\address{Department of Physics, University of Zanjan, University Blvd, 45371-38791, Zanjan, Iran}

%%%%%%%%%%%%%%%%%%%%%%%%%%%%%%%%%%%%%%%%%%%%%%%%%%%%%%%%%%%%%%%%%%
\begin{abstract}
The atom-photon entanglement using the Laguerre-Gaussian beams is studied in the closed-loop three-level $V$-type quantum systems. We consider two schemes with degenerated and non-degenerated upper levels: in the first, the effect of the quantum interference due to the spontaneous emission is taken into account and in the second, a microwave plane wave is applied to the upper levels transition for non-degenerated scheme. It is shown that the atom-photon entanglement in both schemes depends on the intensity profile as well as the orbital angular momentum (OAM) of applied fields so that the various spatially dependent entanglement patterns can be generated by Laguerre-Gaussian beams with different OAMs. However, no entanglement appears in the center of optical vortex beams, because of the zero intensity. As a result, the entanglement between atoms and its spontaneous emissions in different points of the atomic cell can be controlled by the OAM of the applied fields. Moreover, our numerical results show that the number of the local maximum degree of entanglement peaks are determined by the OAM of the applied fields. It seems that the present results  would greatly facilitate the determination of the OAM and may find the broad applications in quantum information processing.
\end{abstract}
%%%%%%%%%%%%%%%%%%%%%%%%%%%%%%%%%%%%%%%%%%%%%%%%%%%%%%%
%%%%%%%%%%%%%%%%%%%%%%%%%%%%%%%%%%%%%%%%%%%%%%%%%%%%%%%
\maketitle
%%%%%%%%%%%%%%%%%%%%%%%%%%%%%%%%%%%%%%%%%%%%%%%%%%%%%%%%%%%%%%%%%%%%

\section{Introduction}
Entanglement is generally a quantum mechanical phenomenon without any classical analogous which makes a correlation between the parts of a multi-partite system \cite{Schrodinger}. A quantum state of a bipartite entangled system cannot be described by a simple tensor product of quantum state of two subsystems \cite{Ficek}. Not only does atom-photon entanglement play an outstanding role in quantum mechanics such as early measurement of Bell inequality violations \cite{Freedman}, but it has also high potential applications in teleporation \cite{tele unknown1993}, cryptography \cite{crypto1991}, error correction \cite{mixed state entan1996} and other quantum computation processes \cite{Theory1}.

Generally speaking, the angular momentum carried by light can be distinguished by the spin angular momentum associated with circular polarization \cite{Beth} and the orbital angular momentum (OAM) associated with the spatial distribution of the wavefront. In 1992, Allen \textit{et al.} showed that LG laser mode has a well-defined OAM, as  $l\hbar$ per photon, and proposed an experiment to observe the torque on cylindrical lenses arising from the reversal of the helicity of a LG mode \cite{Allen}. An additional motivation for recent studies of LG beams is as a route to obtaining narrower the electromagnetically induced transparency (EIT) spectrum than the conventional Gaussian one \cite{Hanle2010,Sapam2011,Akin2014}. The transmission of structured light has been measured using the phase profile in cold rubidium atoms and been shown that the EIT is spatially dependent for vortex light beams \cite{Radwell}. Recently, Mahmoudi \textit{et al.} has investigated the effect of LG intensity profile on the optical spectrum of multi-photon resonance phenomena. It has been found that the linewidth of the optical spectrum due to the multi-photon transition becomes narrower in the presence of a LG beam \cite{Kazemi2016}. They have also studied trap split in an atomic system interacting with  femtosecond LG laser pulses \cite{Kazemi2017}. More recently, we have demonstrated that the LG beam can decrease the full width at the half maximum of the output probe intensity in the electromagnetically induced focusing \cite{Amini}.

It is well known that a LG light beam carries quantized OAM with an azimuthal phase dependence of $\exp(-il\phi)$\cite{padgett}. Inasmuch as, in the closed-loop quantum system the scattering of the coupling fields into the probe field mode is happened \cite{Mahmoudi2006}, the optical properties of such system depend on the relative phase of applied fields. Similarly, it seems that the effect of the OAM of the LG beams can be seen in this system due to the multi-photon transitions.

In this paper, we study the effect of intensity profile on the quantum entanglement in the three-level quantum systems. In first scheme, we consider $V$-type atomic system with two degenerated excited levels. The presence of the quantum coherence due to the spontaneous emission prepares a channel to scattering of the applied fields and makes the system as a closed-loop quantum system. In second scheme, we consider two non-degenerated excited states and apply a microwave field to these states. Then, the optical properties of such systems depend on the OAM of the applied fields. We are interested in studying the entanglement of atom and its spontaneous emission, using the LG fields with different OAMs. We use the von Nuemann entropy to calculating the degree of entanglement (DEM). It is shown that the DEM completely depends on the space and the various points in the atomic cell experience different DEM. Moreover, we find that the DEM is dependent on the both magnitude and sign of the OAM of light beams. Then, the DEM can be controlled by the intensity profile, as well as the OAM of the applied fields. It is worth noting that the maximal DEM is obtained in the second scheme. Because of the high dimensionality of the OAM, the results of present work can be used in a wide variety of systems that use entanglement for quantum information processing. In addition, it suggests a new method    for measuring the OAM of light beams.
%----------------------------------------------------------------------------------

\section{Model and equations}

In general, an entangled quantum system consisting of two subsystems, $A$ and $B$, is described by the reduced density matrix. The state of the entangled system is not a simple tensor product of the subsystems reduced density matrices, $\rho_{AB}\neq\rho_{A}\otimes\rho_{B}$. The von Nuemann entropy, as a helpful quantity for calculating the DEM between the subsystems, is given by
\begin{equation}\label{e1}
    S=-Tr(\rho ln\rho),
\end{equation}
where $\rho$ is the density matrix operator. It can be easily found that the von Neumann entropy vanishes for a bipartite system in pure state and isolated from its environment \cite{1988}. In 1970, Araki and Lieb demonstrated that the subsystems entropies satisfy in the triangle inequality \cite{1970}
\begin{equation}\label{e2}
    |S_{A}(t)-S_{B}(t)|\leq S_{AB}(t)\leq|S_{A}(t)+S_{B}(t)|.
\end{equation}
Here, $S_{AB}(t)$ is the total entropy of the bipartite system which remains constant, for a pure state system, as time proceeds. Using Eqs. \ref{e1} and \ref{e2}, the DEM is obtained as
\begin{equation}\label{e3}
    DEM(t)=S_{A}=S_{B}=-\sum_{i=1}^{3}\lambda_{i}ln\lambda_{i},
\end{equation}
where $\lambda_{i}$ is the reduced density matrix operator eigenvalues. The maximum value of DEM for a $N$-level quantum system is given by $ln N$, whenever the population is uniformly distributed in the dressed states of the system \cite{1991,comment1991,abazari}. Now, we are going to study the effect of LG applied fields on the atom-photon entanglement in two schemes of closed-loop three-level atomic systems.

\subsection{Three-level $V$-type atomic system}

\begin{figure}[htbp]
\centering
  \includegraphics[width=7.0cm]{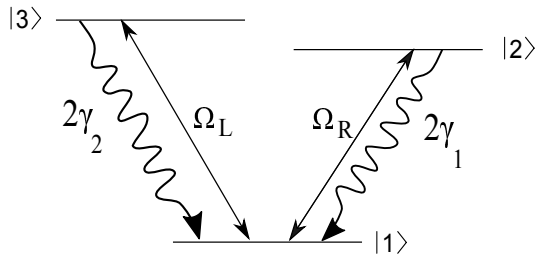}
  \caption{Schematics of the three-level $V$-type atomic system}\label{f1}
\end{figure}

We consider an ensemble of three-level $V$-type atomic system with two near degenerated excited states $|2\rangle$ and $|3\rangle$ and a ground state $|1\rangle$, as shown in Fig. \ref{f1}. As a realistic example, we consider the sodium $D_{2}$ transition with energy levels $|3 ^{2}S_{1/2}\rangle$, $|3 ^{2}P_{1/2}\rangle$ and $|3 ^{2}P_{3/2}\rangle$ corresponding to $|1\rangle$, $|2\rangle$ and $|3\rangle$, respectively. The transition $|1\rangle\leftrightarrow|2\rangle$ is driven by the right LG field with frequency $\omega_{R}$ and Rabi frequency $\Omega_{R}=\vec{\mu}_{12}.\vec{E}_{R}/|hbar$, While the transition $|1\rangle\leftrightarrow|3\rangle$ is excited by the left LG field with frequency $\omega_{L}$ and Rabi frequency $\Omega_{R}=\vec{\mu}_{13}.\vec{E}_{L}/\hbar$. Here $\vec{E}_{R(L)}$ denotes the amplitude of right (left) LG field. The parameter $\mu_{ij}$ represents the induced dipole moment of the $|i\rangle\leftrightarrow|j\rangle$ transition. The spontaneous emissions from two degenerated excited states $|2\rangle$ and $|3\rangle$ to the ground state is denoted by $2\gamma_{1}$ and $2\gamma_{2}$, respectively. In such a system, the spontaneous emission passing through the indistinguishable pathes leads to the quantum interference called as spontaneously generated coherence (SGC).
The interaction Hamiltonian of the system under the electric-dipole and rotating-wave approximations can be written as
\begin{eqnarray}\label{e4}
    H_{I}=-\hbar\Omega_{R}e^{i\Delta_{R}t}|2\rangle\langle1|-\hbar\Omega_{L}e^{i\Delta_{L}t}|3\rangle\langle1|+h.c.,
\end{eqnarray}
where $\Delta_{R}=\omega_{R}-\omega_{21}$ and $\Delta_{L}=\omega_{L}-\omega_{31}$ are the applied fields frequeny detunings with respect to the atomic transition frequencies. The parameter $\omega_{ij}$ is the frequency of $|i\rangle\leftrightarrow|j\rangle$ transition. Both applied field are chosen as the LG fields given by
\begin{eqnarray}\label{e5}
    E_{L}(r,\varphi)&=&E_{0_{L}}\frac{w_{G_{L}}}{\sqrt{|l_{L}|!}w_{LG_{L}}}(\frac{\sqrt{2}r}{w_{LG_{L}}})^{|l_{L}|}e^{-r^{2}/w_{LG_{L}}^{2}}e^{il_{L}\varphi},\nonumber\\
    E_{R}(r,\varphi)&=&E_{0_{R}}\frac{w_{G_{R}}}{\sqrt{|l_{R}|!}w_{LG_{R}}}(\frac{\sqrt{2}r}{w_{LG_{R}}})^{|l_{R}|}e^{-r^{2}/w_{LG_{R}}^{2}}e^{il_{R}\varphi},
\end{eqnarray}
where $E_{0_{L}}$($E_{0_{R}}$), $w_{G_{L}}$($w_{G_{R}}$), $w_{LG_{L}}$($w_{LG_{R}}$) and $l_{L}$($l_{R}$) are the left(right) field amplitude, Gaussian beam waist, LG beam waist and optical angular momentum, respectively. Using the von Nuemann equation, the density matrix equations of motion for such a system takes the form
\begin{eqnarray}\label{e6}
  \dot{\rho}_{22}&=&-2\gamma_{1}\rho_{22}+i\Omega^{\ast}_{R}e^{-i\Delta t}\rho_{12}-i\Omega_{R}e^{i\Delta t}\rho_{21}\nonumber\\
  &-&\eta\sqrt{\gamma_{2}\gamma_{1}}(\rho_{32}+\rho_{23}),\nonumber\\
  \dot{\rho}_{33}&=&-2\gamma_{2}\rho_{33}+i\Omega^{\ast}_{L}\rho_{13}-i\Omega_{L}\rho_{31}
  -\eta\sqrt{\gamma_{2}\gamma_{1}}(\rho_{32}+\rho_{23}),\nonumber\\
  \dot{\rho}_{12}&=&(-\gamma_{1}+i(\Delta_{R}+\Delta))\rho_{12}+i\Omega_{L}\rho_{32}+i\Omega_{R}e^{i\Delta t}(\rho_{22}-\rho_{11})\nonumber\\
  &-&\eta\sqrt{\gamma_{2}\gamma_{1}}\rho_{13},\nonumber\\
  \dot{\rho}_{13}&=&(-\gamma_{2}+i\Delta_{L})\rho_{13}+i\Omega_{R}e^{i\Delta t}\rho_{23}+i\Omega_{L}(\rho_{33}-\rho_{11})\nonumber\\
  &-&\eta\sqrt{\gamma_{2}\gamma_{1}}\rho_{12},\nonumber\\
  \dot{\rho}_{23}&=&(-(\gamma_{1}+\gamma_{2})+i(\Delta_{L}-\Delta_{R}-\Delta))\rho_{23}-i\Omega_{L}\rho_{21}\nonumber\\
  &+&i\Omega^{\ast}_{R}e^{-i\Delta t}\rho_{13}-\eta\sqrt{\gamma_{2}\gamma_{1}}(\rho_{22}+\rho_{33}),\nonumber\\
  \dot{\rho}_{11}&=&-( \dot{\rho}_{22}+ \dot{\rho}_{33}),
\end{eqnarray}
where $\Delta=\Delta_{R}-\Delta_{L}=\omega_{R}-\omega_{L}$. The parameter $\eta$  in Eq. (\ref{e6}) stands for the SGC strength.

 As mentioned before, the dressed states of an atomic system can help us for realization the entanglement behavior which are equal to eigenstates of interaction Hamiltonian. For the proposed model, the dressed states population are given by
\begin{eqnarray}\label{e7}
\rho_{\alpha\alpha}&=&\dfrac{1}{1+B}(\rho_{22}-\sqrt{B}(\rho_{23}+\rho_{32})+\rho_{33}),\nonumber\\
\rho_{\beta\beta}&=&\dfrac{1}{1+A+B}(A\rho_{11}+B\rho_{22}-C(\rho_{12}+\rho_{21})-D(\rho_{13}+\rho_{31})\nonumber\\
&+&\sqrt{B}(\rho_{23}+\rho_{32})+\rho_{33}),\nonumber\\
\rho_{\gamma\gamma}&=&\dfrac{1}{1+A+B}(A\rho_{11}+B\rho_{22}+C(\rho_{12}+\rho_{21})+D(\rho_{13}+\rho_{31})\nonumber\\
&+&\sqrt{B}(\rho_{23}+\rho_{32})+\rho_{33}),
\end{eqnarray}
where
\begin{eqnarray*}
  A &=& \frac{\Omega_{R}^{2}+\Omega_{L}^{2}}{\Omega_{L}^{2}}, \qquad B=\frac{\Omega_{R}^{2}}{\Omega_{L}^{2}},\nonumber\\
  C &=& \frac{\Omega_{R}\sqrt{\Omega_{R}^{2}+\Omega_{L}^{2}}}{\Omega_{L}^{2}}, D=\frac{\sqrt{\Omega_{R}^{2}+\Omega_{L}^{2}}}{\Omega_{L}}.
\end{eqnarray*}
Thus, for physical justification of the quantum entanglement phenomenon, we can study the dressed states population distribution.

\subsection{Closed-loop three-level $V$-type atomic system}

Here, we consider an ensemble of three-level $V$-type atomic system with two non-degenerated excited states and apply a planar microwave field, $\Omega_{m}$,  to the $|2\rangle\leftrightarrow|3\rangle$ transition. The effect of the SGC is dropped in this system. Two other applied fields are same as in previous scheme.
\begin{figure}[htbp]
\centering
  \includegraphics[width=7.0cm]{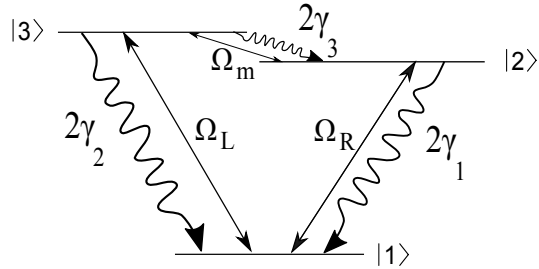}
  \caption{Schematics of the closed-loop three-level $V$-type atomic system}\label{f2}
\end{figure}
Using the von Nuemann equation and the interaction Hamiltonian of the system, the density matrix equations of motion are obtained by
\begin{eqnarray}\label{e8}
  \dot{\rho}_{22}&=&-2\gamma_{1}\rho_{22}+2\gamma_{3}\rho_{33}+i\Omega^{\ast}_{R}\rho_{12}-i\Omega_{R}\rho_{21}+i\Omega_{m}\rho_{32}\nonumber\\
  &-&i\Omega^{\ast}_{m}\rho_{23},\nonumber\\
  \dot{\rho}_{33}&=&-2(\gamma_{2}+\gamma_{3})\rho_{33}+i\Omega^{\ast}_{L}e^{-i\Delta t}\rho_{13}-i\Omega_{L}e^{i\Delta t}\rho_{31}-i\Omega_{m}\rho_{32}\nonumber\\
  &+&i\Omega^{\ast}_{m}\rho_{23}\nonumber\\
  \dot{\rho}_{12}&=&(-\gamma_{1}+i\Delta_{R})\rho_{12}+i\Omega_{L}e^{i \Delta t}\rho_{32}+i\Omega_{R}(\rho_{22}-\rho_{11})\nonumber\\
  &-&i\Omega^{\ast}_{m}\rho_{13},\nonumber\\
  \dot{\rho}_{13}&=&(-(\gamma_{2}+\gamma_{3})+i(\Delta_{L}-\Delta))\rho_{13}+i\Omega_{R}\rho_{23}\nonumber\\
  &+&i\Omega_{L}e^{i \Delta t}(\rho_{33}-\rho_{11})-i\Omega_{m}\rho_{12},\nonumber\\
  \dot{\rho}_{23}&=&(-(\gamma_{1}+\gamma_{2}+\gamma_{3})+i\Delta_{m})\rho_{23}-i\Omega_{L}e^{i \Delta t}\rho_{21}+i\Omega^{\ast}_{R}\rho_{13}\nonumber\\
  &+&i\Omega_{m}(\rho_{33}-\rho_{22}),\nonumber\\
  \dot{\rho}_{11}&=&-( \dot{\rho}_{22}+ \dot{\rho}_{33}),
\end{eqnarray}
where $\Delta=\Delta_{L}-\Delta_{R}-\Delta_{m}$.

\section{Results and discussions}

\subsection{Three-level $V$-type atomic system}

\begin{figure}[htbp]
\centering
  \includegraphics[width=6.0cm]{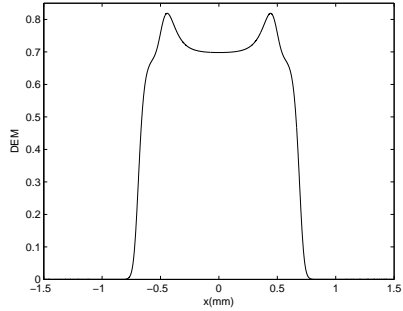}
  \caption{Steady-state behavior of the DEM versus x for $l_{L}=l_{R}=0$. Used parameters are $\gamma_{1}=\gamma_{2}=\gamma=1$, $\eta=0.99$, $\Omega_{0_{L}}=7\gamma$, $\Omega_{0_{R}}=9\gamma$, $w_{G_{L}}=w_{G_{R}}=1mm$, $w_{LG_{L}}=w_{LG_{R}}=270\mu m$ and $\Delta_{L}=\Delta_{R}=0$, multi-photon resonance condition.}\label{f3}
\end{figure}

\begin{figure}[htbp]
\centering
  \includegraphics[width=6.0cm]{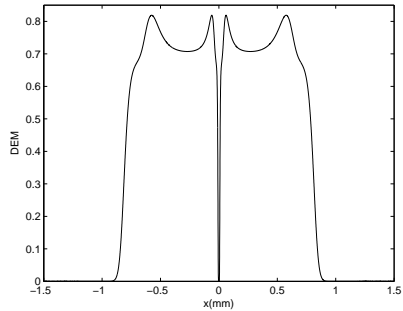}
  \caption{Steady-state behavior of the DEM as a function of x for $l_{L}=l_{R}=2$. Other parameters are same as in Fig. \ref{f3}.}\label{f4}
\end{figure}

Now, we are interested in studying the DEM  for $V$-type atomic system by numerically solving Eqs. \ref{e3} and \ref{e6}. It is assumed that the multi-photon resonance condition is fulfilled. All frequency parameters are scaled by $\gamma_{1}$, which is in the order of $MHz$ for the chosen atomic system. We consider different intensity profiles, i.e., Gaussian and LG for applied fields. Firstly, we assume that the two applied fields have Gaussian modes and the DEM is investigated for different points of atomic cell, in the presence of SGC. Figure \ref{f3} shows the DEM as a function of $x$ for the Gaussian intensity profile of applied fields, i.e., $l_{L}=l_{R}=0$. Used parameters are $\gamma_{1}=\gamma_{2}=\gamma=1$, $\eta=0.99$, $\Omega_{0_{L}}=7\gamma$, $\Omega_{0_{R}}=9\gamma$, $w_{G_{L}}=w_{G_{R}}=1mm$, $w_{LG_{L}}=w_{LG_{R}}=270\mu m$ and $\Delta_{L}=\Delta_{R}=0$. It can be easily seen that the DEM changes from zero to a maximum value, for different values of $x$. Secondly, we consider the LG modes for applied fields and investigate the effect of the intensity profile on the DEM. In Fig. \ref{f4}, we display the DEM for $l_{L}=l_{R}=2$. The zero intensity in the center of LG profile implies  a disentanglement in the center of LG mode.

\begin{figure}[htbp]
\centering
  \includegraphics[width=6.0cm]{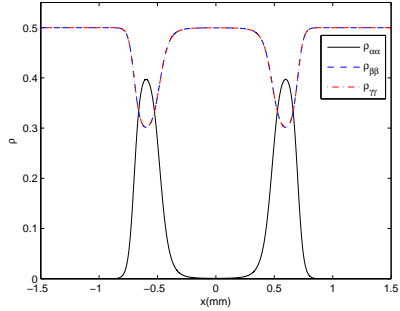}
  \caption{Steady-state population of dressed states versus x for $l_{L}=l_{R}=0$. Other parameters are same as in Fig. \ref{f3}.}\label{f5}
\end{figure}

\begin{figure}[htbp]
\centering
  \includegraphics[width=6.0cm]{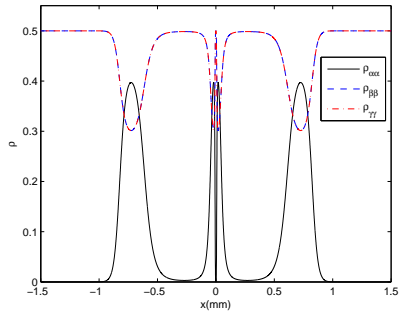}
  \caption{Steady-state population of dressed states as a function of x for $l_{L}=l_{R}=2$. Other parameters are same as in Fig. \ref{f3}.}\label{f6}
\end{figure}

A physical meaning of entanglement lies on the dressed states conception. The uniform population distribution of the dressed states can create the maximal entanglement in atomic system. Using Eq. (\ref{e7}) and numerically solving Eq. (\ref{e6}) the dressed states population can be calculated as a function of $x$. We would like to investigate the dressed states population in different points of the atomic cell. In the following, figures \ref{f5} and \ref{f6} show the dressed states population versus $x$ corresponding to Figs. \ref{f3} and \ref{f4}, respectively.


It is clearly seen that in the maximum entanglement points,  the population is  uniformly distributed in the dressed states as seen in Fig. \ref{f6}.

\begin{figure}[htbp]
\centering
  \includegraphics[width=6.0cm]{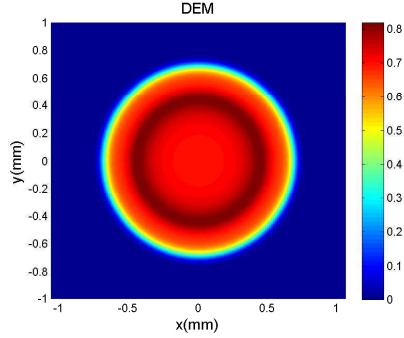}
  \caption{Density plot of the DEM for Gaussian applied fields, $l_{L}=l_{R}=0$. Other parameters are same as in Fig. \ref{f3}.}\label{f7}
\end{figure}

Note that in one dimension, we studied only the effect of the intensity profile on the DEM. Afterwards, we are going to study the effect of OAM of light beams on the atom-photon entanglement in the interested system. So, we have to continue our calculations in two dimensions. As seen in Eq. (\ref{e5}), the effect of the OAM of light appears as a phase term, $e^{il\varphi}$. It was shown that the optical properties of a closed-loop quantum system are dependent on the relative phase of applied fields. Then, it is expected that the optical properties of the $V$-type atomic system depend on the OAM. In Fig. \ref{f7}, we depict the DEM density plot as a function of $x$ and $y$ for $l_{L}=l_{R}=0$. Used parameters are same as Fig. \ref{f3}. It is shown that the local maximum entanglement is occurred in a special radius, in which both applied fields profile are assumed the Gaussian.

\begin{figure}[htbp]
\centering
  \includegraphics[width=4.0cm]{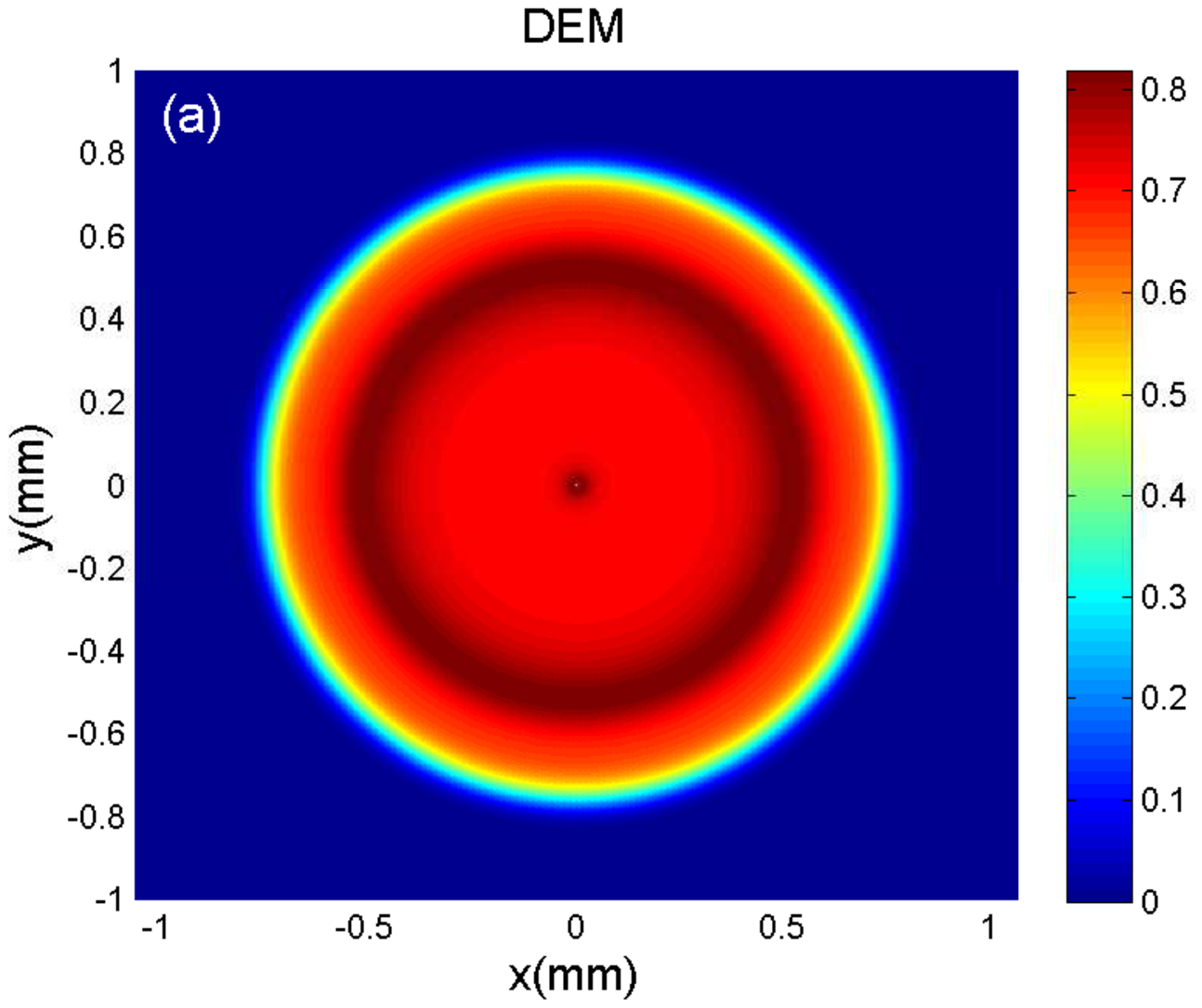} \includegraphics[width=4.0cm]{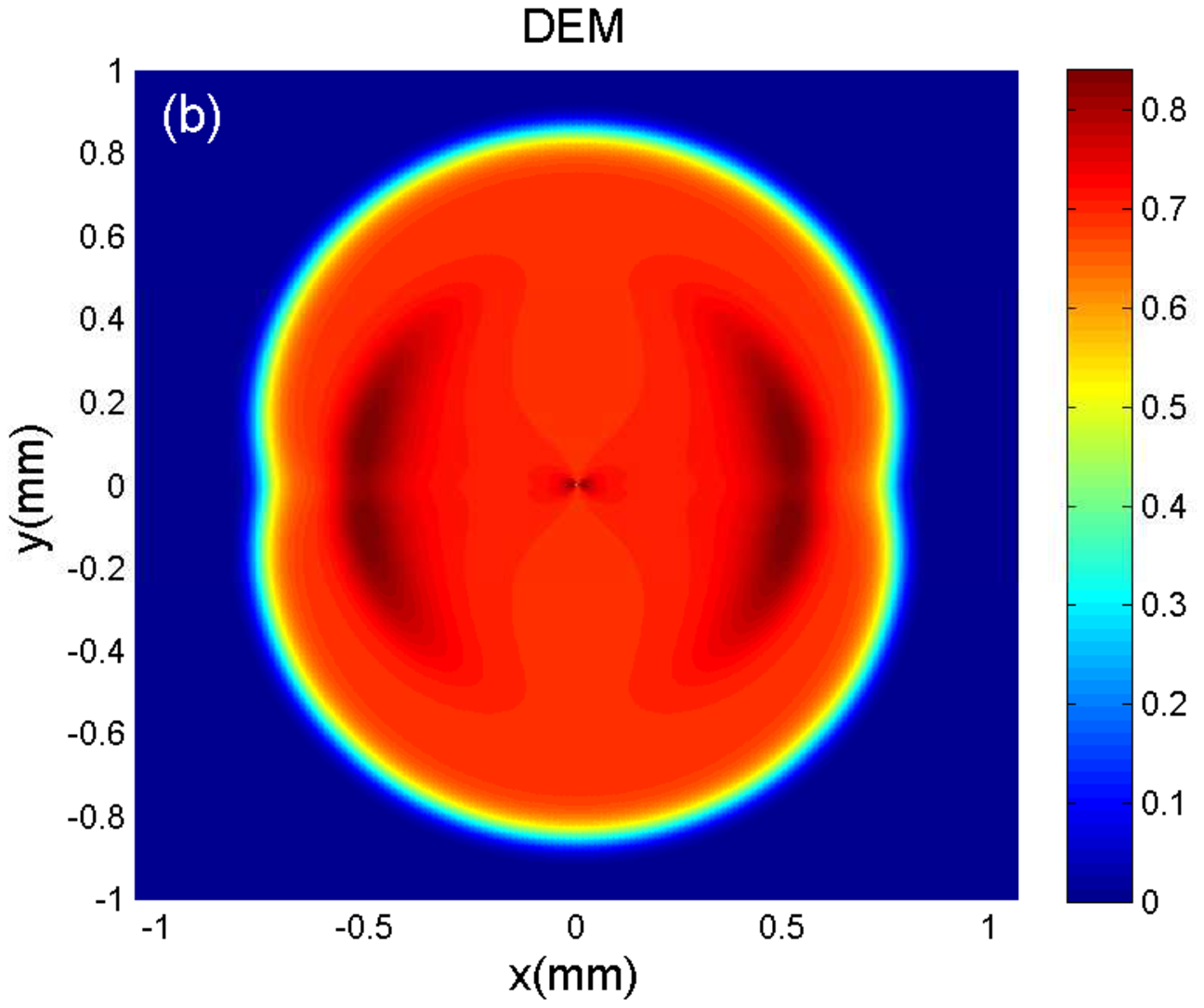}
  \caption{Density plot of the DEM for (a) $l_{L}=l_{R}=1$ and (b) $l_{L}=-l_{R}=1$. Other parameters are same as in Fig. \ref{f2}.}\label{f8}
\end{figure}
In the following, we are interested in studying the effect of the OAM of light beams on the DEM, by considering the LG mode for the applied fields. Figure \ref{f8}(a) shows the DEM density plot for the first mode of LG applied fields, $l_{L}=l_{R}=1$. Other parameters are same as in Fig. \ref{f3}. The DEM value has a central symmetry with a disentanglement point at the origin coordinates due to the zero intensity of applied fields. In figure \ref{f8}(b), we display the DEM density plot, when the OAM of one of the applied fields switches to $l_{R}=-1$. Our results show that the DEM behavior depends on the OAM of applied fields. So, there is an effective way to find the information about the OAM of light beams. The local maximum DEM ring switches to two maximum DEM regions when the rotation direction of wavefront is turned to counter-clockwise. The obtained results are in good agreement with the presented results in references \cite{yao2011,Vickers} in which, it was shown that the superposition pattern of two LG beams with $l$ of opposite sign has a symmetric structure with $|l_{1}|+|l_{2}|$ petals. The DEM profile is similar to the interference pattern of two LG applied fields.

\begin{figure}[htbp]
\centering
  \includegraphics[width=4.0cm]{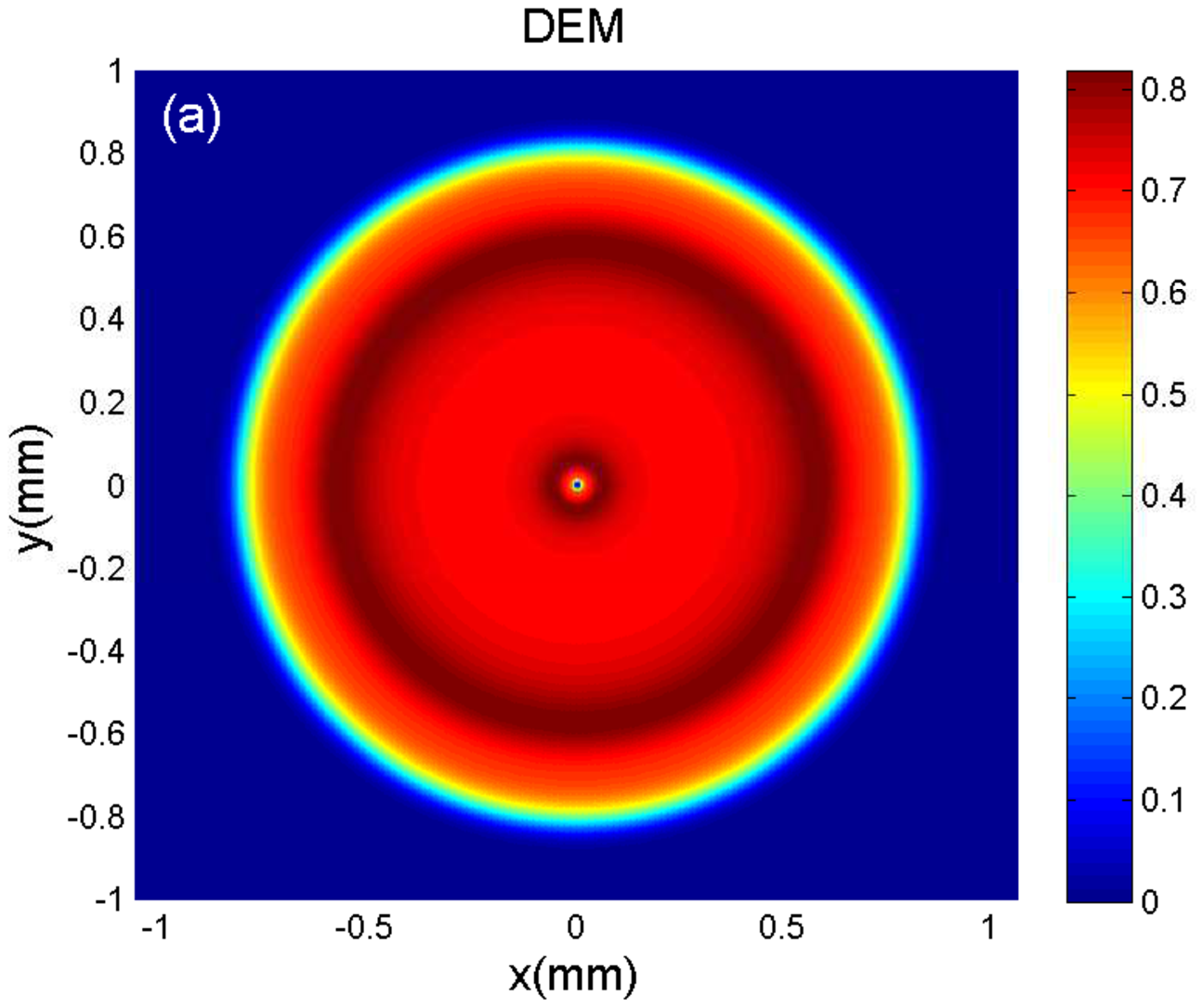} \includegraphics[width=4.0cm]{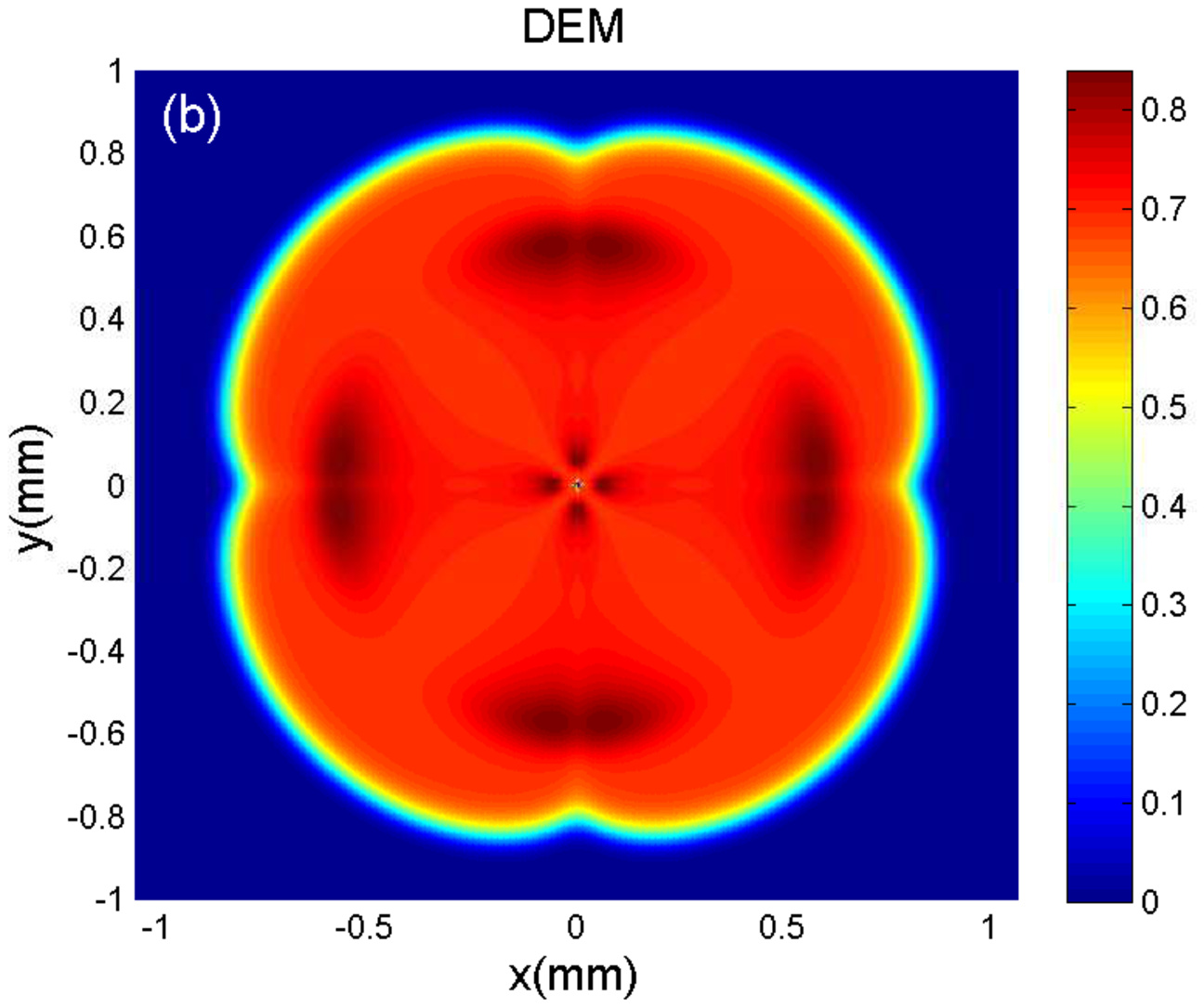}
  \caption{Density plot of the DEM for (a) $l_{L}=l_{R}=2$ and (b) $l_{L}=-l_{R}=2$. Other parameters are same as in Fig. \ref{f3}.}\label{f9}
\end{figure}

 In addition, the DEM density plot for $l_{L}=l_{R}=2$  is shown in Figure \ref{f9}(a). Other used parameters are same as in Fig. \ref{f3}. Two circular areas for maximum DEM can be seen and it is accompanied by an explicit disentanglement at the origin of coordinates. An alternation in the OAM sign of one of applied fields leads to split up the maximum DEM rings to four segments, as shown in Fig. \ref{f9}(b). So, the atom-photon entanglement can be used for realization and determination of OAM of light beams.

\begin{figure}[htbp]
\centering
  \includegraphics[width=6.0cm]{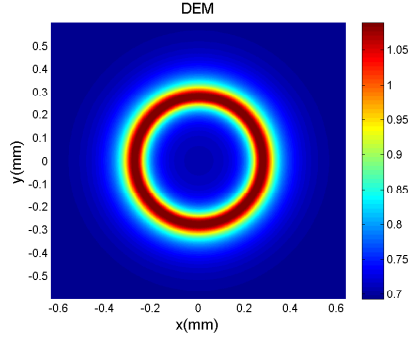}
  \caption{Density plot of the DEM for $l_{L}=0$, $\Omega_{R}=9\gamma$, $\Omega_{L_{0}}=7\gamma$ and beyond multi-photon resonance condition, $\Delta=\gamma$. Other parameters are same as in Fig. \ref{f3}.}\label{f10}
\end{figure}

\begin{figure}[htbp]
\centering
  \includegraphics[width=4.0cm]{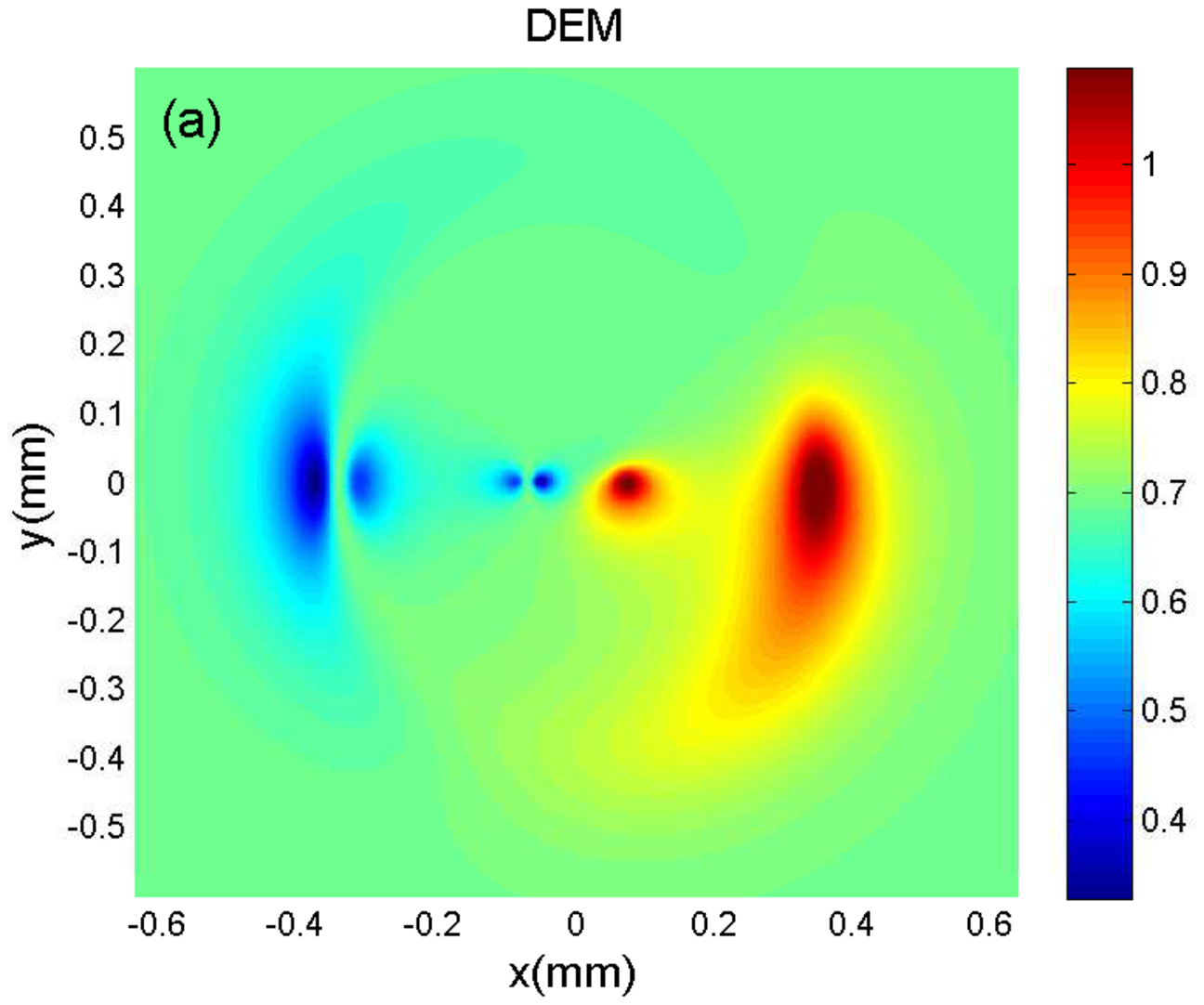} \includegraphics[width=4.0cm]{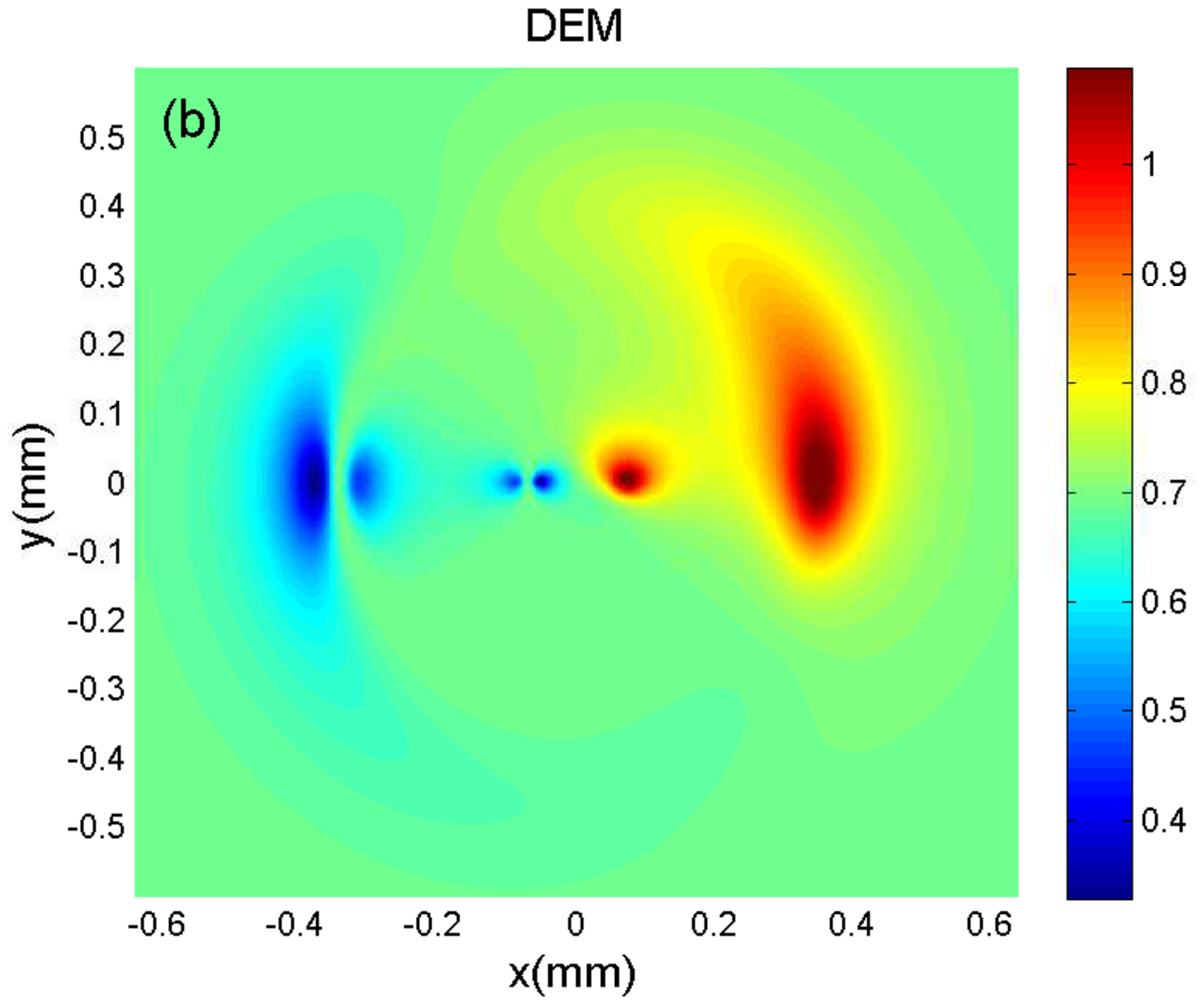}
  \caption{Density plot of the DEM for (a) $l_{L}=1$ and (b) $l_{L}=-1$ under the same parameters as in Fig. \ref{f10}.}\label{f11}
\end{figure}

\begin{figure}[htbp]
\centering
  \includegraphics[width=4.0cm]{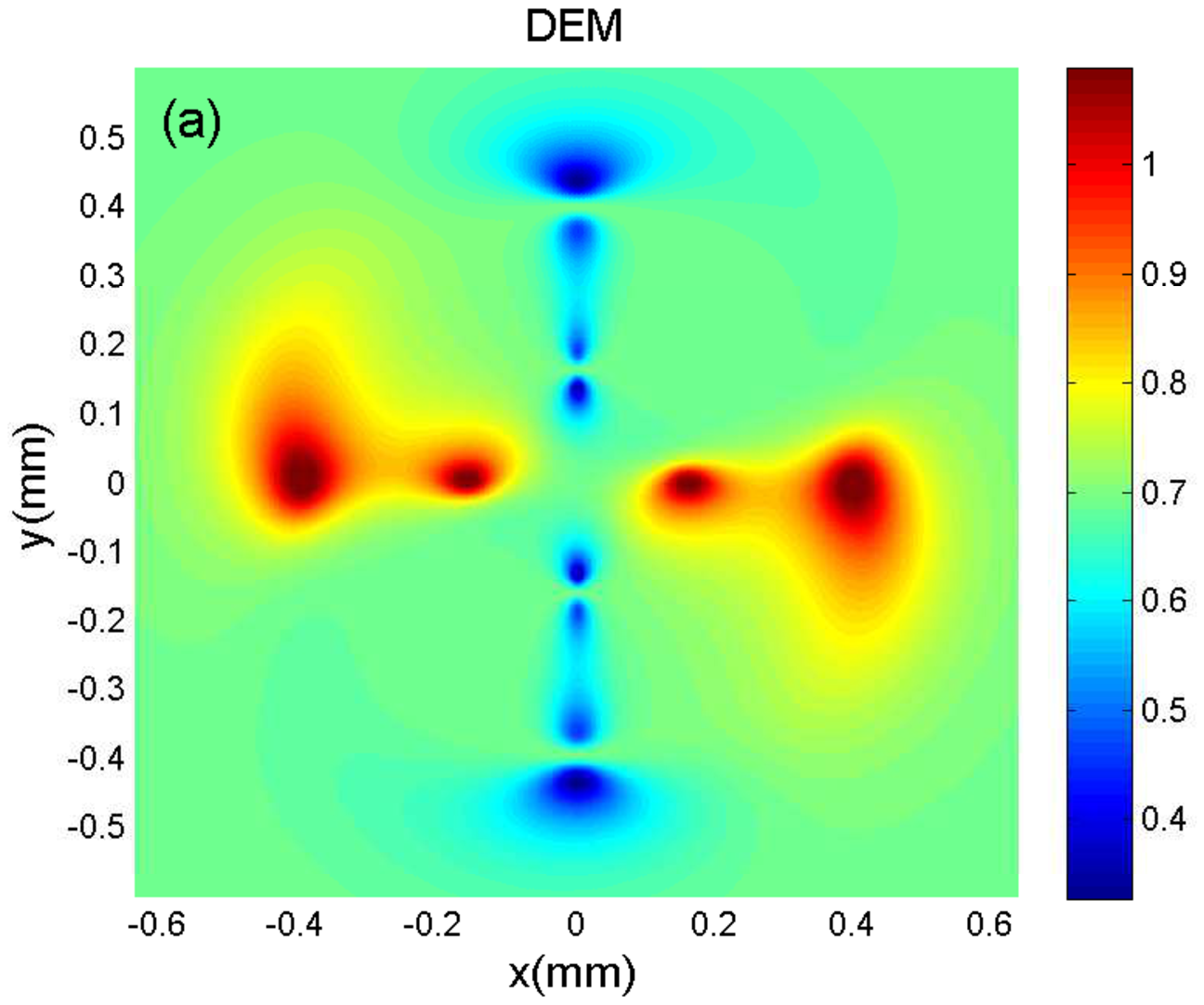} \includegraphics[width=4.0cm]{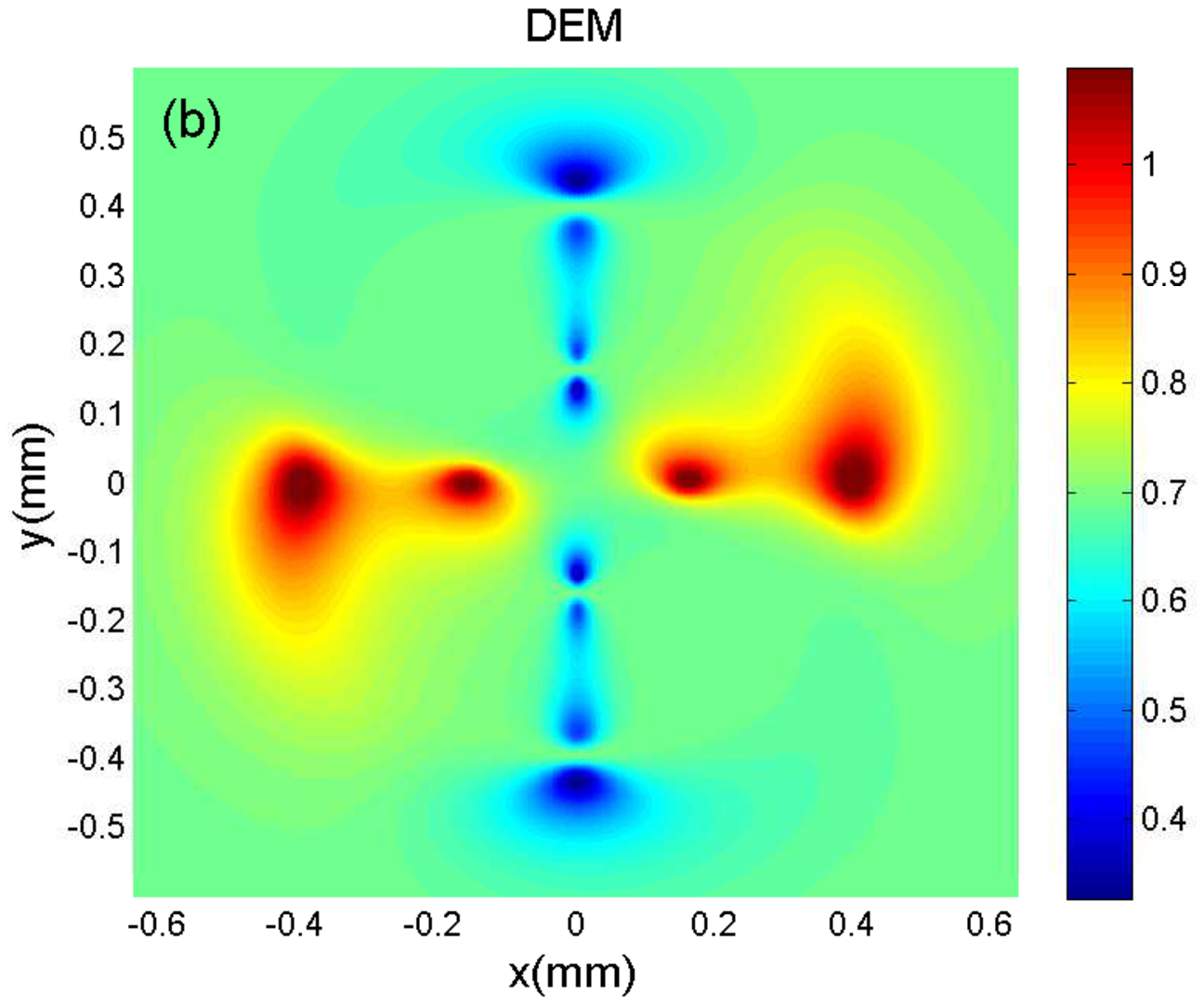}
  \caption{Density plot of the DEM for (a) $l_{L}=2$ and (b) $l_{L}=-2$. Other parameters are same as in Fig. \ref{f10}.}\label{f12}
\end{figure}

 Now, we are going to investigate the DEM behavior of the system beyond multi-photon resonance condition, $\Delta\neq0$. However, we drop the explicit time dependent phase factors in Eq. (\ref{e6}). Then, our results are valid only near multi-photon resonance condition. The right (left) field is considered as a plane-wave (LG field). In Fig. \ref{f10}, we display the DEM density plot as a function of $x$ and $y$ for a Gaussian left field, $l_{L}=0$. Used parameters are chosen as $\Omega_{R}=9\gamma$, $\Omega_{L_{0}}=7\gamma$, $\Delta=\gamma$ and other parameters are same as in Fig. \ref{f3}. It is shown that the DEM nearly reaches to its maximal value, $ln3$, in a special radius. The DEM density plot for the first mode of LG left field is plotted in Fig. \ref{f11}(a) versus $x$ and $y$. Other parameters are same as in Fig. \ref{f10}. The DEM behavior is very similar to spiral-shaped fringes due to the interference of first mode of LG light beam with a plan-wave \cite{padgett,Vickers}. Figure \ref{f11}(b) shows the DEM density plot as a function of $x$ and $y$ for LG left field with $l=-1$ under the same parameters as in Fig. \ref{f10}. It is clearly seen that the DEM behavior for the negative OAM of the first mode of LG left field is in good agreement with the rotation direction of spiral arms of LG field and plane-wave interferogram. We repeat our numerical calculations for the second mode of LG left field in Figs. \ref{f12}(a) and \ref{f12}(b) for $l_{L}=2$ and $l_{L}=-2$, respectively. Other parameters are same as in Fig. \ref{f10}. An investigation on Figs. \ref{f12}(a) and \ref{f12}(b) shows that the DEM density plot is similar to the interference pattern of the second mode of LG beam with the palne wave. Moreover, the both magnitude and sign of the OAM of the applied field can affect the spatially dependent DEM.

\subsection{Closed-loop three-level $V$-type atomic system}

\begin{figure}[htbp]
\centering
  \includegraphics[width=6.0cm]{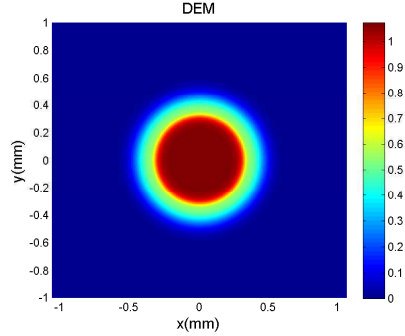}
  \caption{Density plot of the DEM for $l_{L}=l_{R}=0$, $\Omega_{m}=7\gamma$, $\gamma_{3}=2\gamma$ and $\Delta_{m}=0$. Other parameters are same as in Fig. \ref{f3}.}\label{f13}
\end{figure}

Here, we are going to show the behavior of the DEM in a closed-loop three-level $V$-type atomic system using Eqs. \ref{e3} and \ref{e8}. A planar microwave field is applied to the $|2\rangle\leftrightarrow|3\rangle$ transition. It is assumed that the multi-photon resonance condition is fulfilled. Figure \ref{f13} presents the density plot of DEM versus $x$ and $y$ for Gaussian left(right) field. The microwave Rabi frequency, relaxation rate between two exited states and detuning of the microwave field frequency and transition frequency are considered to be $\Omega_{m}=7\gamma$, $\gamma_{3}=2\gamma$ and $\Delta_{m}=0$, respectively. Other used parameter are same as in Fig. \ref{f3}. It is shown that the maximal entanglement can be obtained in a central circular region of atomic cell by applying a planar microwave field instead of the effect of the SGC.

\begin{figure}[htbp]
\centering
  \includegraphics[width=4.0cm]{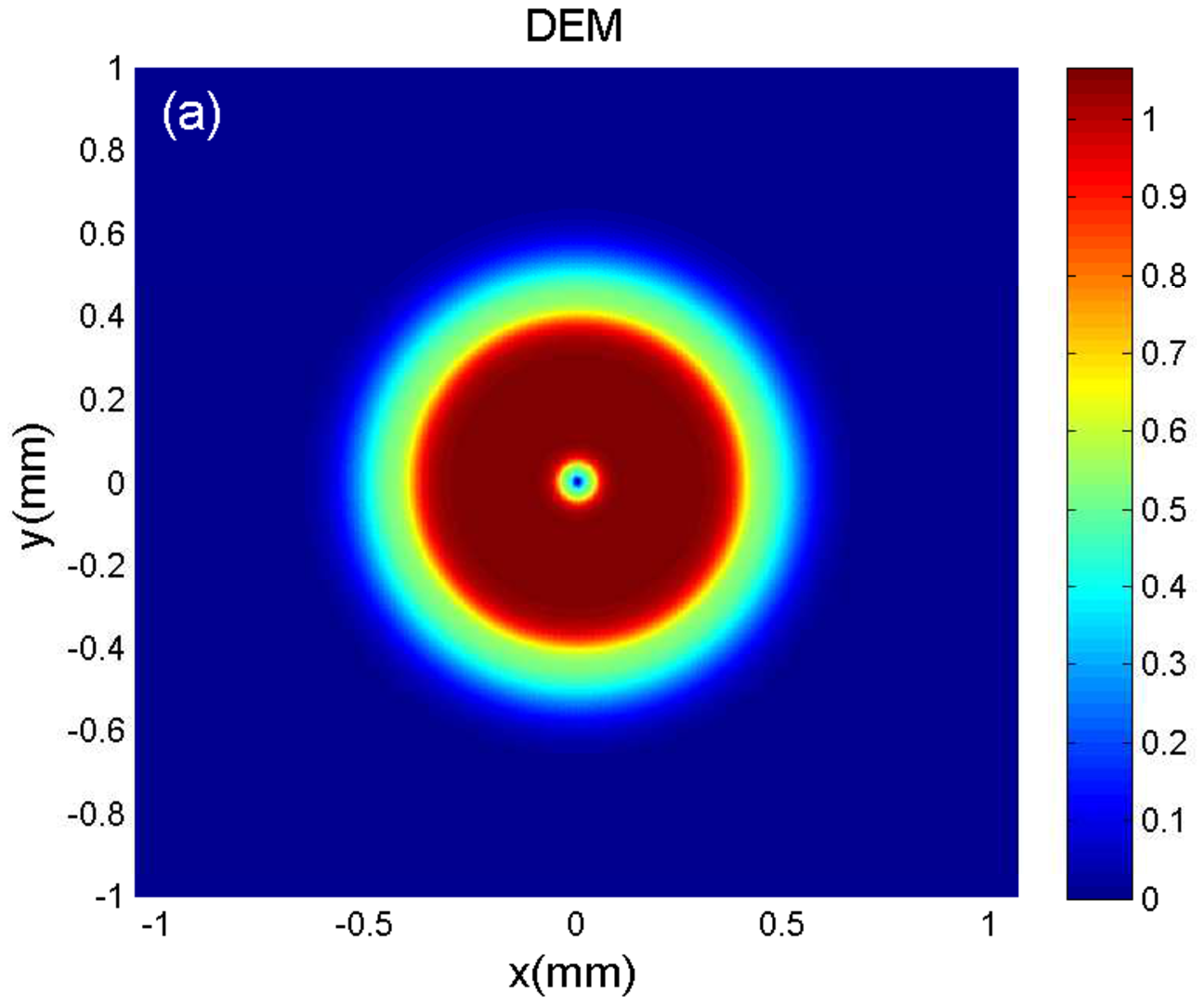} \includegraphics[width=4.0cm]{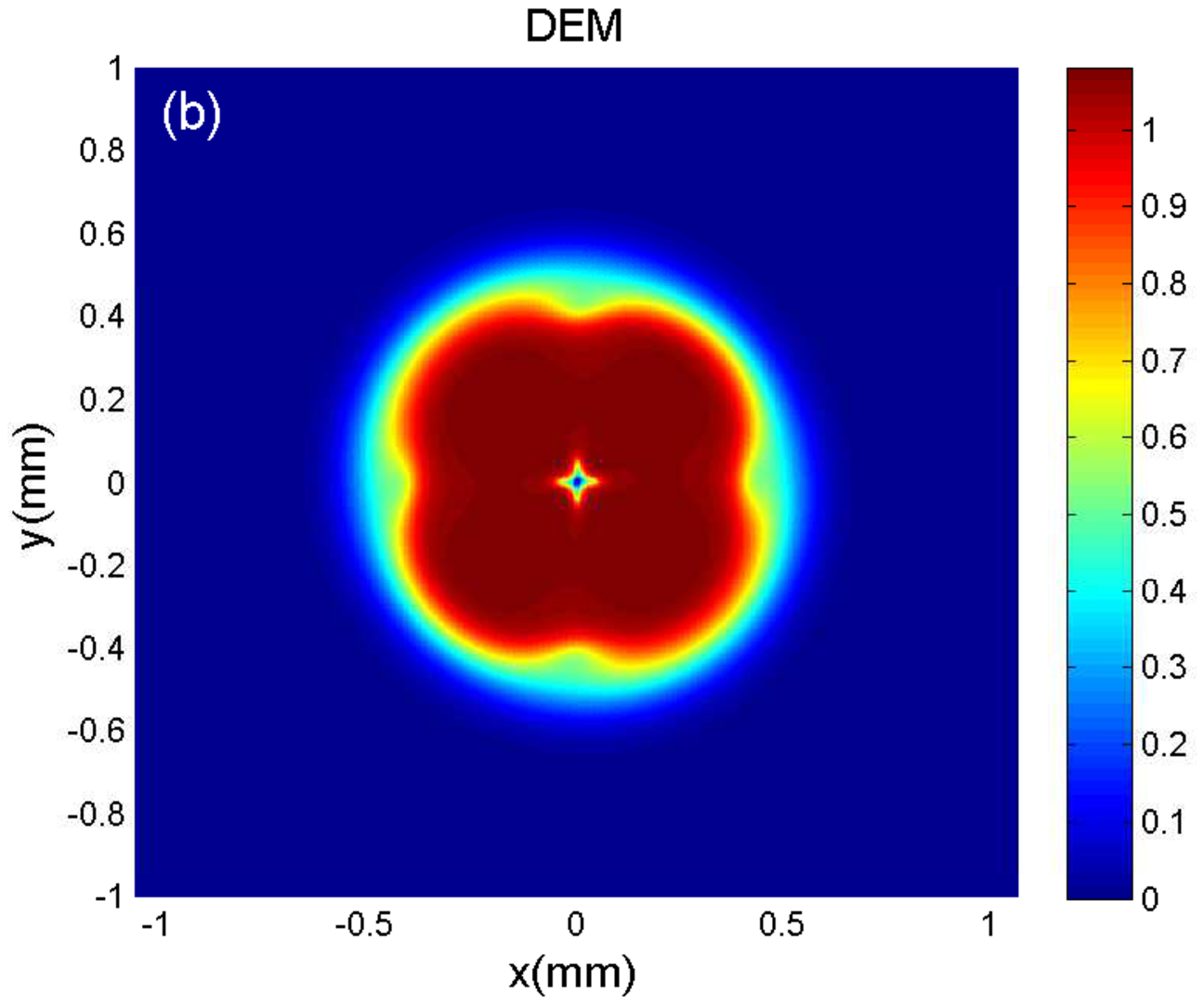}
  \caption{Density plot of the DEM for (a) $l_{L}=l_{R}=1$, (b) $l_{L}=-l_{R}=1$ under the same parameters as in Fig. \ref{f13}.}\label{f14}
\end{figure}

We then proceed to study the effect of OAM of the applied fields on the DEM behavior in the closed-loop $V$-type system. In Fig. \ref{f14}(a), the DEM density plot is depicted as a function of $x$ and $y$ for first mode of LG applied fields with equal OAMs, $l_{L}=l_{R}=1$ for the same parameters of Fig. \ref{f13}. A comparison between Figs. \ref{f8}(a) and \ref{f14}(a) shows that the region of maximal entanglement has been grown and the DEM approximately reaches to its maximal value. The density plot of DEM for two LG fields with equal and opposite OAMs, $l_{L}=1=-l_{R}=1$, is shown in Fig. \ref{f14}(b). An investigation on Figs. \ref{f8}(b) and \ref{f14}(b) shows the different behaviors for the DEM. The maximal entanglement is nearly splitted to four symmetric segments by applying the planar microwave field, $\Omega_{m}$.
\begin{figure}[htbp]
\centering
  \includegraphics[width=4.0cm]{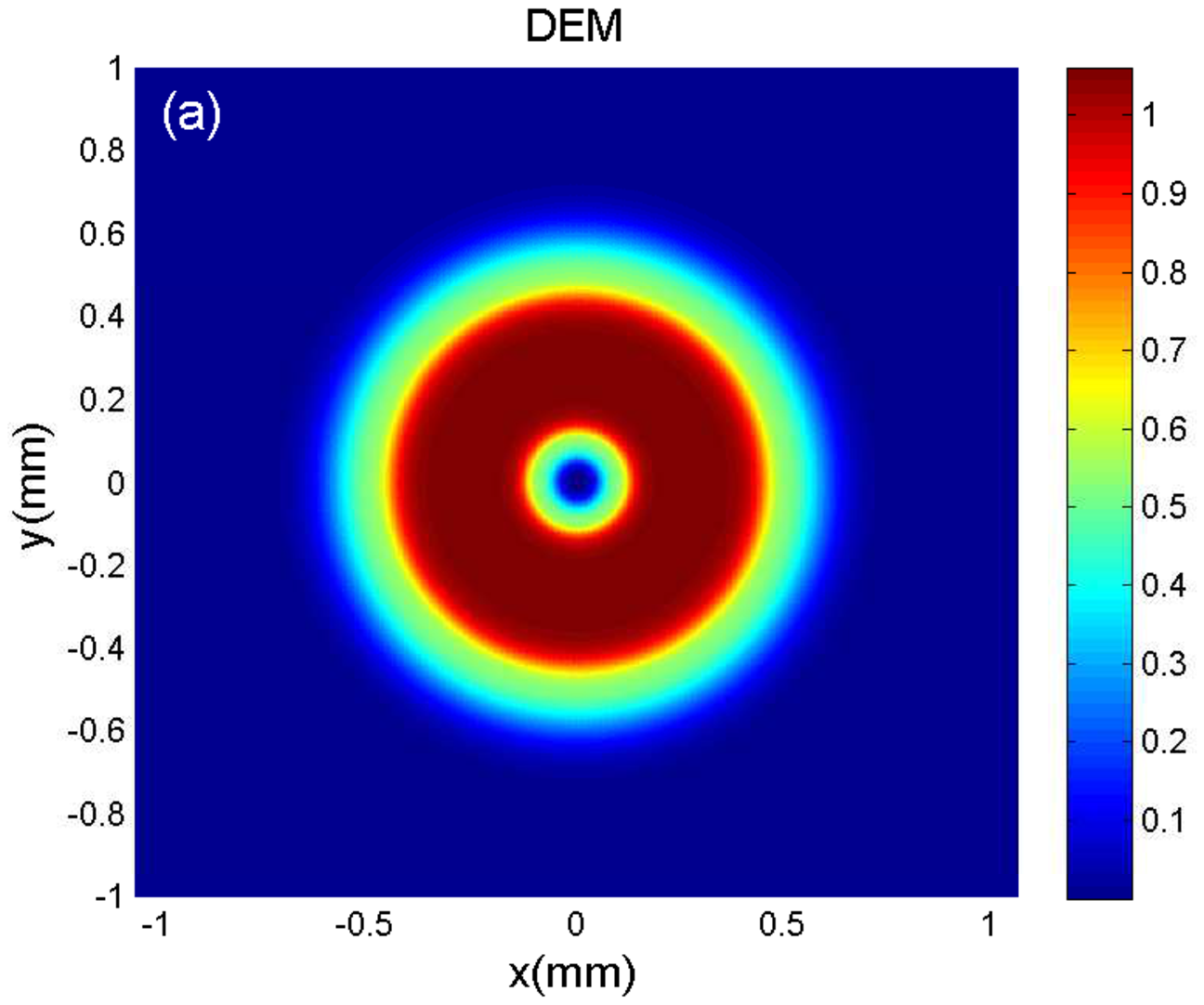} \includegraphics[width=4.0cm]{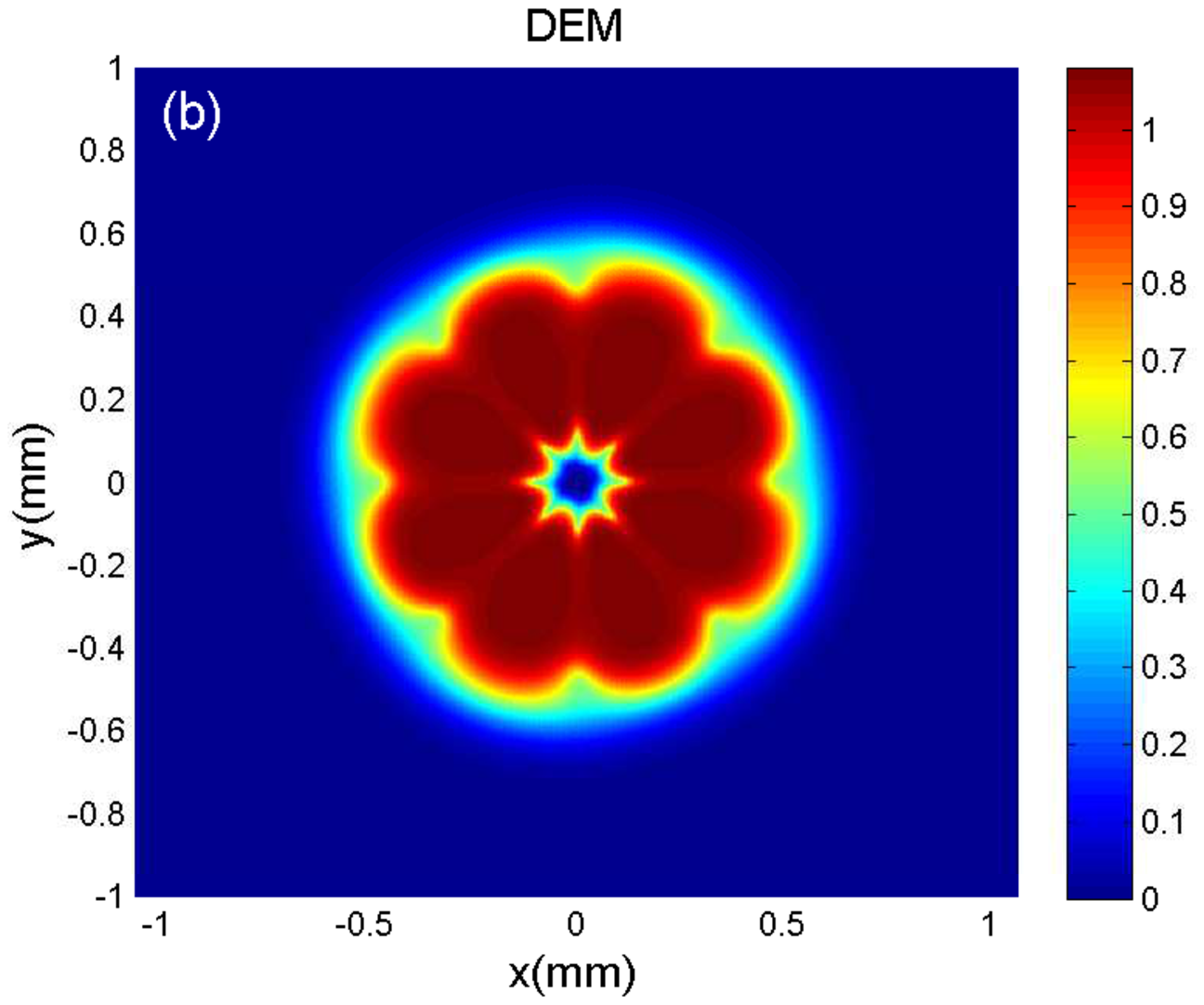}
  \caption{Density plot of the DEM for (a) $l_{L}=l_{R}=2$, (b) $l_{L}=-l_{R}=2$. Other parameters are same as in Fig. \ref{f13}.}\label{f15}
\end{figure}
We plot Fig. \ref{f15}(a) and Fig. \ref{f15}(b) by performing similar calculations for (a) $l_{L}=l_{R}=2$, (b) $l_{L}=-l_{R}=2$. Other parameters are same as in Fig. \ref{f13}. It is demonstrated that the point of central disentanglement are growing up for the second mode of LG fields. Furthermore, the maximal entanglement region split to eight symmetric segments for two LG fields with equal and opposite OAMs, $l_{L}=-l_{R}=2$.

\section{Conclusion}

To conclude, we have studied the quantum entanglement between an ensemble of three-level atomic system and its spontaneous emissions. At first, we investigated the steady-state behavior of the DEM  in a three-level $V$-type atomic system with the SGC effect under the multi-photon resonance condition. We found an intriguing result which illustrates an explicit dependency between the DEM density plot and the OAM of LG light beams. Moreover, we obtained that the DEM density plot pattern is corresponded to the interference pattern of two applied fields. In second scheme, a microwave plane field was applied to the non-degenerated upper levels transition which leads to the maximal DEM. It was demonstrated that the number of maximal entanglement peaks is determined by the OAM of the LG applied fields. The disentanglement and maximal DEM were simultaneously obtained in second scheme. The presented results can be used to determination of the OAM, optical communications and transfer of quantum information.
%----------------------------------------------------------------------------------

%-----------------------------------------------------------------------------------

\begin{thebibliography}{b}
\bibitem{Schrodinger}
E. Schr\"{o}dinger, "Die gegenw\"{a}rtige Situtation in der Quantenmechanik," die Naturwissenschaften \textbf{23}, 807 (1935)

\bibitem{Ficek} Z. Ficek and S. Swain, "Quantum Coherence and Interference; Theory and Experiments," Springer, Berlin (2004)

\bibitem{Freedman} S. J. Freedman and J. F. Clauser, "Experimental test of local hidden variables theories," Phys. Rev. Lett. \textbf{28}, 938-941(1972)

\bibitem{tele unknown1993} C. H. Bennett, G. Brassard, C. Cr\'{e}peau, R. Jozsa, A. Peres and W. K. Wootters, "Teleporting an unknown quantum state via dual classical and Einstein-Podolsky-Rosen channels," Phys. Rev. Lett. \textbf{70}, 1895 (1993)

\bibitem{crypto1991} A. K. Ekert, "Quantum cryptography based on Bell's theorem," Phys. Rev. Lett. \textbf{67}, 661 (1991)

\bibitem{mixed state entan1996} C. H. Bennett, D. P. DiVincenzo, J. A. Smolin and W. K. Wootters, "Mixed-state entanglement and quantum error correction," Phys. Rev. A \textbf{54}, 3824 (1996)

\bibitem{Theory1} G. Benenti, G. Casati and G. Strini, "Principles of Quantum Computation and Information; vol 1: Basic Concepts," World Scientific, Singapore (2004)

\bibitem{Beth} R. A. Beth, "Mechanical detection and measurement of the angular momentum of light," Phys. Rev. \textbf{50}, 115 (1936)

\bibitem{Allen} L. Allen, M. W. Beijersbergen, R. J. C. Spreeuw and J. P. Woerdman, "Orbital angular momentum of light and the transformation of Laguerre-Gaussian laser modes" Phys. Rev. A \textbf{45}, 8185 (1992)

\bibitem{Hanle2010} J. Anupriya, N. Ram, and M. Pattabiraman, "Hanle electromagnetically induced transparency and absorption resonances with a Laguerre Gaussian beam," Phys. Rev. A \textbf{81}, 043804 (2010)

\bibitem{Sapam2011} S. R. Chanu, A. K. Singh, B. Brun, K. Pandey, and V. Natarajan, "Subnatural linewidth in a strongly-driven degenerate two-level system," Optics Communications \textbf{284}, 4957-4960 (2011)

\bibitem{Akin2014} T. G. Akin, S. P. Krzyzewski, A. M. Marino, and E. R. I. Abraham, "Electromagnetically induced transparency with Laguerre-Gaussian modes in ultracold rubidium," Optics Communications \textbf{339}, 209-215 (2015)

\bibitem{Radwell} N. Radwell, T. W. Clark, B. Piccirillo, S. M. Barnett, and S. Franke-Arnold, "Spatially dependent electromagnetically induced transparency," Phys. Rev. Lett. \textbf{114}, 123603 (2015)

\bibitem{Kazemi2016} S. H. Kazemi and M. Mahmoudi, "Multi-photon resonance phenomena using Laguerre-Gaussian beams," Journal of Physics B: Atomic, Molecular and Optical Physics \textbf{49}, 245401 (2016)

\bibitem{Kazemi2017} S. H. Kazemi, S. Ghanbari, and M. Mahmoudi, "Trap split with Laguerre-Gaussian beams," J. Opt. \textbf{19}, 085503 (2017)

\bibitem{Amini} Z. Amini Sabegh, M. A. Maleki, and M. Mahmoudi, "Phase-controlled electromagnetically induced focusing in a closed-loop atomic system," J. Opt. Soc. Am. B \textbf{34}, 2446-2451 (2017)

\bibitem{padgett} M. Padgett, J. Courtial, and L. Allen, "Light's orbital angular momentum," Physics Today \textbf{57}, 35-40 (2004)

\bibitem{Mahmoudi2006} M. Mahmoudi and J. Evers, "Light propagation through closed-loop atomic media beyond the multiphoton resonance condition," Phys. Rev. A \textbf{74}, 063827 (2006)

\bibitem{1988} S. J. D. Phoenix and P. L. Knight, "Fluctuations and entropy in models of quantum optical resonance," Ann. Phys. \textbf{186}, 381-407 (1988)

\bibitem{1970} H. Araki and E. H. Lieb, "Entropy inequalities," Commun. Math. Phys. \textbf{18}, 160-170 (1970)

\bibitem{1991} S. J. D. Phoenix and P. L. Knight, "Establishment of an entangled atom-field state in the Jaynes-Cummings model," Phys. Rev. A \textbf{44}, 6023 (1991)

\bibitem{comment1991} S. J. D. Phoenix and P. L. Knight, "Comment on "Collapse and revival of the state vector in the Jaynes-Cummings model: An example of state preparation by a quantum apparatus"," Phys. Rev. Lett. \textbf{66}, 2833 (1991)

\bibitem{abazari} M. Abazari, A. Mortezapour, M. Mahmoudi, and M. Sahrai, "Phase-controlled atom-photon entanglement in a three-level $V$-type atomic system via spontaneously generated coherence," Entropy \textbf{13}, 1541-1554 (2011)

\bibitem{yao2011} A. M. Yao and M. J. Padgett, "Orbital angular momentum: origins, behavior and applications," Advances in Optics and Photonics \textbf{3}, 161-204 (2011)

\bibitem{Vickers} J. Vickers, M. Burch, R. Vyas and S. Singh, "Phase and interference properties of optical vortex beams," J. Opt. Soc. Am. A \textbf{25}, 823-827 (2008)
\end{thebibliography}
\end{document}